\documentclass[journal]{IEEEtran}
\usepackage{amsmath,amsfonts,amssymb}
\usepackage[dvips]{graphicx}
\usepackage{verbatim}
\usepackage{slashbox}
\usepackage{xcolor}
\usepackage{subfigure}
\usepackage{bm}
\usepackage{algorithm} 
\usepackage{algorithmic} 
\usepackage{balance}
\usepackage[T1]{fontenc}
\usepackage{cite}
\usepackage{changepage}
\IEEEoverridecommandlockouts

\newcommand{\figwidth}{9.2}

\begin{document}

\title{Achievable Sum Rates of Half- and Full-Duplex Bidirectional OFDM Communication Links}

\author{Zhenyu Xiao,~\IEEEmembership{Member,~IEEE,}
        Yao Li,
        Lin Bai,~\IEEEmembership{Member,~IEEE,}
        Jinho Choi,~\IEEEmembership{Senior Member,~IEEE}

\thanks{This work was partially supported by the National Natural Science Foundation of China (NSFC) under grant Nos. 61571025, 61201189, 91338106, 91538204, and 61231013, and Foundation for Innovative Research Groups of the National Natural Science Foundation of China under grant No. 61221061.}
\thanks{Z. Xiao, Y. Li and L. Bai are are with the School of
Electronic and Information Engineering, Beijing Key Laboratory for Network-based Cooperative Air Traffic Management, and Beijing Laboratory for General Aviation Technology, Beihang University, Beijing 100191, P.R. China.}
\thanks{J. Choi is with the School of Information and Communications, Gwangju Institute of Science and Technology (GIST), Korea.}

\thanks{Corresponding Author: Dr. L. Bai with Email: l.bai@buaa.edu.cn.}
}

\maketitle
\begin{abstract}
While full-duplex (FD) transmission has the potential to double the system capacity, its substantial benefit can be offset
by the self-interference (SI) and non-ideality of practical transceivers.
In this paper, we investigate the achievable sum rates (ASRs)
of half-duplex (HD) and FD transmissions with orthogonal frequency division multiplexing (OFDM), where the non-ideality is taken into consideration. Four transmission strategies are considered, namely HD with uniform power allocation (UPA), HD with non-UPA (NUPA), FD with UPA, and FD with NUPA.
For each of the four transmission strategies,
an optimization problem is formulated to maximize its ASR,
and a (suboptimal/optimal) solution with low complexity is accordingly
derived.
Performance evaluations and comparisons are conducted
for three typical channels, namely symmetric frequency-flat/selective and asymmetric frequency-selective channels. Results show that the proposed solutions for both HD and FD transmissions can achieve near
optimal performances. For FD transmissions, the optimal solution can be obtained under typical conditions. In addition,
several observations are made
on the ASR performances of HD and FD transmissions.
\end{abstract}


\section{Introduction}
Full-duplex wireless communication (FDWC), which allows simultaneous
transmission and reception
(Tx/Rx) in the same frequency band \cite{Sabharwal2014JSAC,ahmed_2012,duarte_2011,duarte_2014,jain_2011}, have attracted
increasing attention recently due to
the potential of doubling the spectrum efficiency.
Recent research has considered and demonstrated the feasibility of FDWC in practice \cite{Sabharwal2014JSAC, ahmed_2012,duarte_2011,duarte_2014,jain_2011}. One of the most critical challenges of FDWC is to cancel the self-interference (SI)
from a local transmitter to a local receiver \cite{Sabharwal2014JSAC,li_2012}.
For example, a typical transmission power of WiFi signals
is about 20 dBm (100 mW), which may lead to significant SI far above the noise floor $-90$ dBm \cite{bharadia_2013}.
Thus, in FDWC, without SI cancellation,
the bit-error-rate (BER) performance of wireless links, which
is determined by the signal-to-interference-plus-noise ratio (SINR),
will dramatically deteriorate.

There are a number of existing works on the SI cancelation. These works basically exploit antenna cancelation, (analog) radio-frequency (RF) cancellation and (digital) baseband cancellation, or a multi-stage approach with the combination of them. For instance,
Duarte et al. proposed a RF cancellation scheme \cite{Sabharwal2014JSAC,duarte_2011,duarte_2014}, where an extra transmit RF chain is used to generate a reference RF signal for SI cancelation in analog domain.
A similar approach was also adopted in \cite{lis_2011}. Jain et al.
proposed a different scheme with balun cancelation in RF \cite{jain_2011}, which uses signal inversion through a balun circuit. Choi et al. proposed an antenna cancellation scheme by making
use of two transmit antennas
with distances $d_1$ and $d_2$ from the receive antenna,
where the distance difference, $|d_1 - d_2|$, is half-wavelength
\cite{choi_2010}. Offsetting the two transmit antennas by
half-wavelength causes for a destructive addition of their signals
and canceling them.
In \cite{khoj_2011}, the architectures of two symmetric transmit antennas with respect to (w.r.t.) the receive antenna
(or two symmetric receive antennas w.r.t. the transmit antenna)
have
been proposed to cancel SI. These results also show that it is hard to cancel
the SI at a sufficient level
using only
antenna cancelation, RF cancelation or baseband cancelation, and a combination of them might be necessary in practice \cite{Sabharwal2014JSAC,duarte_2011,duarte_2014,jain_2011,choi_2010,khoj_2011}. A recent work in \cite{bharadia_2013} shows that a full duplex radio with the combination approach is able to cancel at a level of 110 dB SI. Moreover, in the regime of multiple-input multiple-output (MIMO) FD systems, precoding and beamforming are also exploited to mitigate SI \cite{huberman2015mimo,huberman2014self,huberman2015self,zhangjs2013robust,cirik2014mse}.

In addition to SI, FDWC also faces another challenge under non-ideal transceiver implementations. For a practical communication link, the non-ideality
comes from a cascaded effect containing both transmitter and receiver imperfections \cite{Wei2014}, such as IQ imbalances, phase noise, non-linearity of amplifiers and so on \cite{ahmed_2012,sahai_2012_unders}. Due to non-ideality, the actual constellation points are deviated from their ideal locations, and this deviation is usually quantified by a measure called error vector magnitude (EVM) level \cite{Wei2014}, which is extensively used in many standards \cite{3Gpp2012}. In FDWC, as the SI is usually much stronger than the received information-bearing signal from a remote transmitter, the EVM noise is significant and cannot be neglected. For tractability,
in this paper, we model the EVM noise as zero-mean Gaussian distributed \cite{Wei2014,Suziki2008,Santella1998,Day2012journal,Day2012trans}.

By taking the SI and transceiver non-ideality into consideration, we study the achievable sum rate (ASR) of FDWC, and compare it with that of half-duplex (HD) transmission in this paper. Since orthogonal frequency division multiplexing (OFDM) is widely adopted in current advanced communications \cite{Derrik2012}, we consider it for the multiplexing scheme in this paper. There are existing related works on the achievable rate of FDWC. In \cite{jia2015optimal,cirik2014achievable,cirik2013weighted}, (weighted) sum rate maximization was studied by optimizing power allocation and precoding matrix in MIMO FD systems, where EVM noise was not considered. In \cite{ahmed2013rate}, the regime in which a practical FD systems can achieve
higher rates than an equivalent HD system was analytically studied, with hardware and implementation imperfections, including low-noise amplifier (LNA) noise figure, quantization noise, and phase noise, taken into account. In \cite{Day2012journal} and \cite{Day2012trans}, the non-ideality effects on the sum rate performances of more general MIMO FD bidirectional and relay links were investigated. In \cite{li2014full}, the rate region of an FD single-carrier system was studied by taking the EVM noise effect into account. However, these works \cite{ahmed2013rate,Day2012journal,Day2012trans,li2014full} are based on the single-carrier modulation. In \cite{Wei2014}, one-end rate regions of HD and FD bidirectional link were investigated with OFDM modulation, where both SI and EVM noise were considered. Based on \cite{Wei2014}, we further consider ASR of the two ends in this paper.

It is noteworthy that the maximization of ASR
needs both time allocation over the two nodes and the energy allocation over the two nodes and all the sub-carriers. The optimization problems are challenging because of the complicated objective functions and
a large number of variables. Thus, they can hardly be solved by the conventional convex-optimization methods. The ASR optimization problems under consideration differ from those in \cite{Wei2014}. Firstly, the objective
is different, i.e., ASR is optimized in this paper, while one-end achievable rate is optimized in \cite{Wei2014}. Secondly,
a sum energy constraint is adopted in this paper, while individual energy
constraints were adopted in \cite{Wei2014}. As a consequence, in \cite{Wei2014} the monotonicity of the one-end achievable rate versus the allocated time and energy can be exploited to find optimal/suboptimal solutions, while in this paper the ASR does not have monotonicity versus the time and energy due to the total time and energy constrains, which makes
the problems difficult. On the other hand, the problems under consideration also differ from conventional time allocation for TDD systems \cite{jeon2000comparison} and power allocation for OFDM systems \cite{wong1999multiuser,shen2003optimal}, because joint time and energy allocation with the effects of EVM and SI is considered in this paper, which makes the problem more challenging and difficult to solve.

The contributions of this paper are two-fold. Firstly,
we formulate optimization problems for the four typical transmission strategies, namely HD with uniform power allocation (UPA), HD with non-UPA (NUPA), FD with UPA, and FD with NUPA, and propose an approach to solve the four optimization problems, which first deals with a partial Lagrangian function without taking into account the inequality constraints and formulate an equivalent problem of the original problem, and then solves the new problem with appropriate numerical methods. The proposed approach greatly lowers the search complexity compared with
the exhaustive grid search method, and turns out
to be effective in solving this
category of optimization problems
with time and energy allocations, especially when the number of variables is large. Additionally, we conduct performance evaluations and comparisons for the four transmission strategies under three typical channel conditions, namely symmetric frequency-flat/selective and asymmetric frequency-selective channels.
Results show that the proposed solutions for both HD and FD transmissions can achieve near
optimal performances. For FD transmissions, the optimal solution can be obtained under typical conditions.
In addition,
several observations are made
on the ASR performances of HD and FD transmissions.

The rest of this paper is organized as follows. In Section \ref{section2}, system and channel models for HD and FD bidirectional single-antenna OFDM links, as well as the EVM noise model, are established. In Section \ref{section3}, optimization problems for HD and FD bidirectional links with UPA and NUPA strategies are formulated and solved, respectively. In Section \ref{section4}, performance evaluations and comparisons are conducted under different channel settings. Section \ref{section5} summarizes the paper.
%

\section{System and Channel Models}
\label{section2}

In this paper, we consider the same system and channel models
as those in \cite{Wei2014}. There are two peer nodes, namely Node 1 and Node 2, with a single antenna in our system, and both point-to-point bidirectional HD and FD OFDM links are considered between these two nodes \cite{Wei2014,Sabharwal2014JSAC,jain_2011,bharadia_2013}. For the HD OFDM link, TDD based resource sharing is adopted to realize bidirectional transmission. While for the FD OFDM link, bidirectional transmission is realized at the same time slot and frequency band. For convenience, in this paper we call the link from Node 1 to Node 2 a forward link, and the link from Node 2 to Node 1 a backward link\footnote{It is noteworthy that a forward (backward) link usually refers to the link from a base station to a mobile station (reverse), but in this paper, a forward (backward) link may also refers to the link from an arbitrary peer node to another (reverse) in a non-infrastructured network.}.

In the OFDM links under consideration,
let $s\left[ k \right]$, $\eta \left[ k \right]$, $x\left[ k \right]$, $h[k]$,
$n[k]$, and $y[k]$ denote the transmitted original signal at the digital
baseband, the EVM noise to capture the transceiver non-ideality, the final transmitted signal, the channel gain, the thermal noise,
and the received signal on the $k$-th sub-carrier, respectively.
Thus, we have $x\left[ k \right] = s\left[ k \right] + \eta \left[ k \right]$,
$k = 1, \ldots, K$, where $K$ is the number of sub-carriers.
Here, we model $\eta \left[ k \right]$ as a zero-mean additive Gaussian noise. Specially, in an FD transceiver chain we add $\eta[k]$ at the receiver as the SI, which will be further discussed later. Define the signal-to-EVM power ratio (SER) as

\begin{equation}
{\gamma _E} = \frac{{\mathbb{E}\left\{ {{{\left| {s\left[ k \right]} \right|}^2}} \right\}}}{{\mathbb{E}\left\{ {{{\left| {\eta \left[ k \right]} \right|}^2}} \right\}}},
\end{equation}
where $\mathbb{E}\{\cdot\}$ represents the expectation operation.

As mentioned, we consider an HD OFDM bidirectional link using a TDD based resource sharing scheme. In this scheme an overall time duration is normalized to be 1, and the forward signals are transmitted from Node 1 to Node 2 during period ${t_1}$, while the backward signals are transmitted from Node 2 to Node 1 during period ${t_2}$, where ${t_1}+t_2=1$. The received signals at two nodes over the $k$-th sub-carrier are respectively given by
\begin{equation}
{y_2}\left[ k \right] = {h_{21}}\left[ k \right]\sqrt {\frac{{{\varepsilon _1}\left[ k \right]}}{{{t_1}}}} \left( {{s_1}\left[ k \right] + {\eta _1}\left[ k \right]} \right) + {n_2}\left[ k \right],
\end{equation}
and
\begin{equation}
{y_1}\left[ k \right] = {h_{12}}\left[ k \right]\sqrt {\frac{{{\varepsilon _2}\left[ k \right]}}{{{t_2}}}} \left( {{s_2}\left[ k \right] + {\eta _2}\left[ k \right]} \right) + {n_1}\left[ k \right],
\end{equation}
where subscript $i$ is the node index, ${h_{ji}}\left[ k \right]$ is the channel coefficient of the link from node $i$ to node $j$, $s_i\left[ k \right]$ is the transmitted signal with a normalized power of 1,
and ${\varepsilon_i}\left[ k \right]$ is the energy
consumed on the $k$-th sub-carrier.

In addition, as in \cite{Wei2014}, we consider an FD OFDM bidirectional link with an overall time period normalized to be
1. The FD system configuration does not change during the whole period. The FD transceivers at the two nodes transmit signals to and receive signals from the other node simultaneously on the same
time slot and carrier frequency. The promising 3-stage SI cancellation is adopted in the FD transceivers, which includes antenna isolation (cancellation), RF cancellation and baseband cancellation. In the antenna isolation and RF cancellation stages, both the SI and EVM noise are mitigated.
We define ${h_{ii}}\left[ k \right]$ as the equivalent channel gain capturing the effect of the remaining SI and EVM \cite{Wei2014}.
In the digital baseband cancellation stage
only the remaining SI can be further reduced to some degree, while the remaining EVM cannot. Hence, we define the attenuate factor ${\beta _i}\left[ k \right]$ to model the function of this stage for
SI cancellation. Accordingly, the received signals at the two nodes over the $k$-th sub-carrier are respectively given by

\begin{equation}
\begin{aligned}
 &{y_2}\left[ k \right] = {h_{21}}\left[ k \right]\sqrt {{\varepsilon _1}\left[ k \right]} \left( {{s_1}\left[ k \right] + {\eta _1}\left[ k \right]} \right)  +\\
 &{h_{22}}\left[ k \right]{\beta _2}\left[ k \right]\sqrt {{\varepsilon _2}\left[ k \right]} {s_2}\left[ k \right] + {h_{22}}\left[ k \right]\sqrt {{\varepsilon _2}\left[ k \right]} {\eta _2}\left[ k \right] + {n_2}\left[ k \right],
\end{aligned}
\end{equation}
and
\begin{equation}
\begin{aligned}
&{y_1}\left[ k \right] = {h_{12}}\left[ k \right]\sqrt {{\varepsilon _2}\left[ k \right]} \left( {{s_2}\left[ k \right] + {\eta _2}\left[ k \right]} \right) + \\
&{h_{11}}\left[ k \right]{\beta _1}\left[ k \right]\sqrt {{\varepsilon _1}\left[ k \right]} {s_1}\left[ k \right]+ {h_{11}}\left[ k \right]\sqrt {{\varepsilon _1}\left[ k \right]} {\eta _1}\left[ k \right] + {n_1}\left[ k \right].
\end{aligned}
\end{equation}

It is noteworthy that
we consider the two nodes involved in the bidirectional link as a whole,
which is different from \cite{Wei2014}.
Therefore, based on the system and channel models, we can optimize the ASR of the HD and FD OFDM bidirectional links with different power allocation strategies over sub-carriers, i.e., UPA and NUPA, under a
total energy constraint.

\section{Optimization of Achievable Sum Rate}
\label{section3}

In this section, we formulate the four optimization problems to maximize the ASR under the total time and energy constraints. The challenges of these problems are: (i) the objective functions are complicated and basically not convex/concave; (ii) there are inequality (non-negative) constraints on the time and energy constraints in addition to the total time and energy constraints; (iii) the number of energy variables is large for the NUPA transmission strategies. To solve these problems, there are two possible candidates in general, namely the exhaustive grid search and the interior-point method \cite{vanderbei1999interior}. The exhaustive grid search directly performs grid search on the independent variables and finds the best value set of the variables. When the step length is set sufficiently small, the performance of the exhaustive grid search can approach the optima, whereas the search complexity will be very high. Additionally, although the interior-point method may be also feasible to obtain a suboptimal solution for a non-convex problem, a linear equation array with more than $M$ variables need to be solved in each iteration \cite{vanderbei1999interior}, where $M$ is the number of original optimization variables. Hence, for the NUPA transmission the complexity of the interior-point method would be also very high.

In this paper, we propose a low-complexity approach to solve these problems\footnote{The optimization problem for the FD-UPA transmission is simple; thus it does not need to adopt the proposed approach.}. In particular, We deal with a partial Lagrangian function without taking into account the inequality constraints first, and then formulate an equivalent problem of the original problem to use the Lagrangian function as the objective function. Finally we solve the new problem by establishing an equation set with the first-order condition of an optima and solving the equation set with low-complexity numerical search methods.


\subsection{ASR of HD Bidirectional Link with UPA}
\subsubsection{Problem Formulation}
In this subsection, we study the maximization of the ASR of an HD bidirectional link with UPA. We assume that the consumed energy on the $k$-th sub-carrier is ${\varepsilon _1}\left[ k \right] = {\varepsilon _1}$ for the forward link and ${\varepsilon _2}\left[ k \right] = {\varepsilon _2}$ for the backward link.
Therefore, the achievable rates of the forward link and backward link, respectively, are
\begin{equation}
\begin{aligned}
{r_1}\left( {{\varepsilon _1},{t_1}} \right) = \frac{{{t_1}}}{K}\sum\limits_{k = 1}^K {{{\log }_2}\left( {1 + \frac{{{\varepsilon _1}{\gamma _E}{{\left| {{h_{21}}\left[ k \right]} \right|}^2}/{t_1}}}{{{\varepsilon _1}{{\left| {{h_{21}}\left[ k \right]} \right|}^2}/{t_1} + \left( {{\gamma _E} + 1} \right){N_2}}}} \right)}
\end{aligned}
\end{equation}
and
\begin{equation}
\begin{aligned}
{r_2}\left( {{\varepsilon _2},{t_2}} \right) = \frac{{{t_2}}}{K}\sum\limits_{k = 1}^K {{{\log }_2}\left( {1 + \frac{{{\varepsilon _2}{\gamma _E}{{\left| {{h_{12}}\left[ k \right]} \right|}^2}/{t_2}}}{{{\varepsilon _2}{{\left| {{h_{12}}\left[ k \right]} \right|}^2}/{t_2} + \left( {{\gamma _E} + 1} \right){N_1}}}} \right)},
\end{aligned}
\end{equation}
where $N_1$ and
$N_2$ are the noise powers on each sub-carrier w.r.t. the power of pure signals excluding the EVM noise for the
forward and backward links, respectively.
Accordingly, the ASR of the HD bidirectional link is given by
\begin{equation}
r\left( {{\varepsilon _1},{\varepsilon _2},{t_1},{t_2}} \right) = {r_1}\left( {{\varepsilon _1},{t_1}} \right)+{r_2}\left( {{\varepsilon _2},{t_2}} \right).
\end{equation}

For convenience, we define
\begin{equation}
\begin{aligned}
{\gamma _E}{\left| {{h_{21}}\left[ k \right]} \right|^2} &= {A_{k1}} > 0,\\
{\left| {{h_{21}}\left[ k \right]} \right|^2} &= {B_{k1}} > 0,\\
\left( {{\gamma _E} + 1} \right){N_2} &= {C_{k1}} > 0,\\
{\gamma _E}{\left| {{h_{12}}\left[ k \right]} \right|^2} &= {A_{k2}} > 0,\\
{\left| {{h_{12}}\left[ k \right]} \right|^2} &= {B_{k2}} > 0,\\
\left( {{\gamma _E} + 1} \right){N_1} &= {C_{k2}} > 0,
\end{aligned}
\end{equation}
and rewrite the sum rate function as
\begin{equation}
\begin{aligned}
r\left( {{\varepsilon _1},{\varepsilon _2},{t_1},{t_2}} \right) = &\frac{{{t_1}}}{K}\sum\limits_{k = 1}^K {{{\log }_2}\left( {1 + \frac{{{A_{k1}}\frac{{{\varepsilon _1}}}{{{t_1}}}}}{{{B_{k1}}\frac{{{\varepsilon _1}}}{{{t_1}}} + {C_{k1}}}}} \right)} \\
&+ \frac{{{t_2}}}{K}\sum\limits_{k = 1}^K {{{\log }_2}\left( {1 + \frac{{{A_{k2}}\frac{{{\varepsilon _2}}}{{{t_2}}}}}{{{B_{k2}}\frac{{{\varepsilon _2}}}{{{t_2}}} + {C_{k2}}}}} \right)} .
\end{aligned}
\end{equation}

To achieve the maximum ASR, we formulate an optimization problem as
\begin{equation} \label{eq_HD-UPA_Problem}
\begin{aligned}
\mathop {{\rm{minimize}}}\limits_{{\varepsilon_1},~{\varepsilon _2},~{t_1},~{t_2}} \quad &-r\left( {{\varepsilon_1},{\varepsilon _2},{t_1},{t_2}} \right),\\
{\rm{subject\ to\quad }}&{t_1} + {t_2} = 1,\\
&K\left( {{\varepsilon _1} + {\varepsilon _2}} \right) = E,\\
&{\varepsilon _1} \ge 0,~{\varepsilon _2} \ge 0,\\
&{t_1} \ge 0, ~{t_2} \ge 0,
\end{aligned}
\end{equation}
where $E$ is the maximum total energy consumed by Node 1 and Node 2, and the fact has been exploited that the maximum ASR is achieved only when all the energy is allocated.

\subsubsection{Solution of the Problem}
As this problem is clearly non-convex and complicated, we are interested in finding a suboptimal solution with three steps.

\emph{Step 1:} Let us first deal with a partial Lagrangian function, i.e., we don't take into account the inequality constraints in the formulation of the Lagrangian function, but we will take care of them when minimizing the partial Lagrangian function. Thus, the partial Lagrangian function is given by
\begin{equation}
\begin{aligned}
&L\left( {{\varepsilon _1},{\varepsilon _2},{t_1},{t_2},\lambda,v} \right) \\
= &- r\left( {{\varepsilon _1},{\varepsilon _2},{t_1},{t_2}} \right) + \lambda \left( {{\varepsilon _1} + {\varepsilon _2} - \frac{E}{K}} \right)+ v\left( {{t_1} + {t_2} - 1} \right),
\end{aligned}
\end{equation}
where $\lambda$ and $v$ are Lagrange multipliers.

\emph{Step 2:} We formulate a new problem, which is equivalent to the original problem in \eqref{eq_HD-UPA_Problem}, to minimize the partial Lagrangian function as follows.
\begin{equation} \label{eq_HD-UPA_Problem_Partial}
\begin{aligned}
\mathop {{\rm{minimize}}}\limits_{{\varepsilon_1},~{\varepsilon _2},~{t_1},~{t_2},\lambda,v} \quad &L\left( {{\varepsilon _1},{\varepsilon _2},{t_1},{t_2},\lambda,v} \right),\\
{\rm{subject\ to\quad }}&{t_1} + {t_2} = 1,\\
&K\left( {{\varepsilon _1} + {\varepsilon _2}} \right) = E,\\
&{\varepsilon _1} \ge 0,~{\varepsilon _2} \ge 0,\\
&{t_1} \ge 0, ~{t_2} \ge 0,
\end{aligned}
\end{equation}

\emph{Step 3:} We solve \eqref{eq_HD-UPA_Problem_Partial} by formulating an equation set and proposing a numerical method to solve the equation set.

At a local optima, we have
\begin{equation}
\begin{aligned}
\frac{{\partial L}}{{\partial {\varepsilon _1}}} = \frac{{\partial L}}{{\partial {\varepsilon _2}}} = \frac{{\partial L}}{{\partial {t_1}}} = \frac{{\partial L}}{{\partial {t_2}}} = 0,\\
\end{aligned}
\label{equ:maxL_HDUA}
\end{equation}

Consequently, we can establish two equations as follows:
\begin{equation} \label{eq_UPA_HD_Eps}
\frac{{\partial L}}{{\partial {\varepsilon _1}}} =  - \frac{{\partial r}}{{\partial {\varepsilon _1}}} + \lambda =
\frac{{\partial L}}{{\partial {\varepsilon _2}}} =  - \frac{{\partial r}}{{\partial {\varepsilon _2}}} + \lambda ,
\end{equation}
\begin{equation} \label{eq_UPA_HD_t}
\frac{{\partial L}}{{\partial {t_1}}} =  - \frac{{\partial r}}{{\partial {t_1}}} + v=
\frac{{\partial L}}{{\partial {t_2}}} =  - \frac{{\partial r}}{{\partial {t_2}}} + v,
\end{equation}
where
\begin{equation}
\frac{{\partial r}}{{\partial {\varepsilon _1}}} = \frac{1}{{K\ln 2}}\sum\limits_{k = 1}^K {\frac{{{A_{k1}}{C_{k1}}}}{{\left[ {\left( {{A_{k1}} + {B_{k1}}} \right)\frac{{{\varepsilon _1}}}{{{t_1}}} + {C_{k1}}} \right]\left[ {{B_{k1}}\frac{{{\varepsilon _1}}}{{{t_1}}} + {C_{k1}}} \right]}}} ,
\end{equation}
\begin{equation}
\frac{{\partial r}}{{\partial {\varepsilon _2}}} = \frac{1}{{K\ln 2}}\sum\limits_{k = 1}^K {\frac{{{A_{k2}}{C_{k2}}}}{{\left[ {\left( {{A_{k2}} + {B_{k2}}} \right)\frac{{{\varepsilon _2}}}{{{t_2}}} + {C_{k2}}} \right]\left[ {{B_{k2}}\frac{{{\varepsilon _2}}}{{{t_2}}} + {C_{k2}}} \right]}}} ,
\end{equation}
\begin{equation}
\begin{aligned}
&\frac{{\partial r}}{{\partial {t_1}}} = \frac{1}{K}\sum\limits_{k = 1}^K {{{\log }_2}\left( {1 + \frac{{{A_{k1}}}}{{{B_{k1}} + {C_{k1}}\frac{{{t_1}}}{{{\varepsilon _1}}}}}} \right)} \\
&- \frac{{{t_1}}}{{{\varepsilon _1}K\ln 2}}\sum\limits_{k = 1}^K {\frac{{{A_{k1}}{C_{k1}}}}{{\left( {{A_{k1}} + {B_{k1}} + {C_{k1}}\frac{{{t_1}}}{{{\varepsilon _1}}}} \right)\left( {{B_{k1}} + {C_{k1}}\frac{{{t_1}}}{{{\varepsilon _1}}}} \right)}}} ,
\end{aligned}
\end{equation}
and
\begin{equation}
\begin{aligned}
&\frac{{\partial r}}{{\partial {t_2}}} = \frac{1}{K}\sum\limits_{k = 1}^K {{{\log }_2}\left( {1 + \frac{{{A_{k2}}}}{{{B_{k2}} + {C_{k2}}\frac{{{t_2}}}{{{\varepsilon _2}}}}}} \right)} \\
&- \frac{{{t_2}}}{{{\varepsilon _2}K\ln 2}}\sum\limits_{k = 1}^K {\frac{{{A_{k2}}{C_{k2}}}}{{\left( {{A_{k2}} + {B_{k2}} + {C_{k2}}\frac{{{t_2}}}{{{\varepsilon _2}}}} \right)\left( {{B_{k2}} + {C_{k2}}\frac{{{t_2}}}{{{\varepsilon _2}}}} \right)}}} .
\end{aligned}
\end{equation}

For convenience, let ${p_1} = \frac{{{\varepsilon _1}}}{{{t_1}}}$ and ${p_2} = \frac{{{\varepsilon _2}}}{{{t_2}}} = \frac{{{E}/{K} - {\varepsilon _1}}}{{1 - {t_1}}}$. Then, \eqref{eq_UPA_HD_Eps} and \eqref{eq_UPA_HD_t} are rewritten as
\begin{equation} \label{eq_HD-UPA_Equation_set}
\begin{aligned}
\left\{ \begin{aligned}
{f_1}\left( {{p_1},{p_2}} \right) = 0,\\
{f_2}\left( {{p_1},{p_2}} \right) = 0,
\end{aligned} \right.
\end{aligned}
\end{equation}
where
\begin{equation}
\begin{aligned}
{f_1}\left( {{p_1},{p_2}} \right) = &\sum\limits_{k = 1}^K {\frac{{{A_{k1}}{C_{k1}}}}{{\left[ {\left( {{A_{k1}} + {B_{k1}}} \right){p_1} + {C_{k1}}} \right]\left[ {{B_{k1}}{p_1} + {C_{k1}}} \right]}}} \\
&- \sum\limits_{k = 1}^K {\frac{{{A_{k2}}{C_{k2}}}}{{\left[ {\left( {{A_{k2}} + {B_{k2}}} \right){p_2} + {C_{k2}}} \right]\left[ {{B_{k2}}{p_2} + {C_{k2}}} \right]}}} ,
\end{aligned}
\end{equation}
and
\begin{equation}
\begin{aligned}
&{f_2}\left( {{p_1},{p_2}} \right) = \\
&\sum\limits_{k = 1}^K {{{\log }_2}\left( {1 + \frac{{{A_{k1}}{p_1}}}{{{B_{k1}}{p_1} + {C_{k1}}}}} \right)} - \sum\limits_{k = 1}^K {{{\log }_2}\left( {1 + \frac{{{A_{k2}}{p_2}}}{{{B_{k2}}{p_2} + {C_{k2}}}}} \right)} \\
&- \frac{1}{{\ln 2}}\left( {\frac{1}{{{p_1}}} - \frac{1}{{{p_2}}}} \right)\sum\limits_{k = 1}^K {\frac{{{A_{k1}}{C_{k1}}}}{{\left[ {\left( {{A_{k1}} + {B_{k1}}} \right){p_1} + {C_{k1}}} \right]\left[ {{B_{k1}}{p_1} + {C_{k1}}} \right]}}},
\end{aligned}
\end{equation}
where ${f_1}\left( {{p_1},{p_2}} \right) =0$ is exploited to simplify the expression of ${f_2}\left( {{p_1},{p_2}} \right)$.

The remaining task is to solve this transcendental equation array, which is difficult and cannot be directly solved. Hence, we choose to use the Newton's method (also known as Newton-Raphson method) with backtracking line search \cite[Chapter 9.2]{boyd2004convex}, which has been shown to be effective and fast to solve both optimization problems \cite[Chapter 9.5]{boyd2004convex} and transcendental equation arrays \cite{Equation_Solving}. For the Newton's method, in each iteration the intersection point of the tangent planes of the functions ($y=f_1(p_1,p_2)$ and $y=f_2(p_1,p_2)$) at the current approximation with the zero plane ($y=0$) is treated as a new approximation \cite{Equation_Solving}. Hence, we need to obtain the descent direction as shown in \eqref{equ:set2_HDUA}. To realize the backtracking line search, we define a function $f(\overline{p})=f_1(p_1,p_2)^2+f_2(p_1,p_2)^2$. The solution to \eqref{eq_HD-UPA_Equation_set} is in fact the minima of $f(p_1,p_2)$. Hence, the step length and the stopping criterion can be defined with $f(p_1,p_2)$. The algorithm is described as Algorithm \ref{alg:LNmethod_HDUA}.

\begin{algorithm}\caption{Newton's Method with Backtracking Line Search to Obtain $p_1$ and $p_2$.}
\label{alg:LNmethod_HDUA}
\begin{algorithmic}
\STATE \textbf{1) Initialize:}
\begin{quote}
--Set initial values $\overline{p}=[p_1,p_2]^{\rm{T}}$ with all the constraints in the original problem satisfied;

--Set a positive error-tolerance variable $\delta  \ll 1$;

--Set two constants $\alpha,~\beta$ with $0<\alpha<0.5,~0<\beta<1$ and step size $t=1$ for the backtracking line search.

\end{quote}

\STATE \textbf{2) Iteration:}
\begin{quote}
\textbf{For} $j=1,2,...,$
\begin{adjustwidth}{0.5 cm}{0 cm}
--Obtain the descent direction: $\Delta \overline{p}=[\Delta p_1,\Delta p_2]^{\rm{T}}$ by solving the linear equation array as shown in \eqref{equ:set2_HDUA};

--Update the step length of backtracking line search: If $f(\overline{p}+t\Delta\overline{p})>f(\overline{p})+{\alpha}t{\Delta}f^{\rm{T}}{\Delta}\overline{p}$, where ${\Delta}f=[\frac{{\partial {f}}}{{\partial {p_1}}},\frac{{\partial {f}}}{{\partial {p_2}}}]^{\rm{T}}$, $t:=\beta t$;

--Update the variables: $\overline{p}:=\overline{p}+t{\Delta}\overline{p}$;

--Stop criterion: if $ f(\overline{p})  < \delta^2 $
    \begin{adjustwidth}{0.5 cm}{0 cm}
    Break;
    \end{adjustwidth}

\end{adjustwidth}
\end{quote}

\STATE \textbf{3) Result:}
\begin{quote}
Return the solution of the equation array $\overline{p}={[p_1,{p_2}]^{\mathop{\rm T}\nolimits} }$.
\end{quote}

\end{algorithmic}
\end{algorithm}

\begin{figure*}[!t]
\begin{equation}
\begin{aligned}
\left\{ \begin{array}{l}
{l_1}\left( \Delta \overline{p} \right):{f_1}\left( {{p_1}^{\left( j \right)},{p_2}^{\left( j \right)}} \right) + \frac{{\partial {f_1}}}{{\partial {p_1}}}\left( {{p_1}^{\left( j \right)},{p_2}^{\left( j \right)}} \right){\Delta p_1} + \frac{{\partial {f_1}}}{{\partial {p_2}}}\left( {{p_1}^{\left( j \right)},{p_2}^{\left( j \right)}} \right){\Delta p_2} = 0\\
{l_2}\left( \Delta \overline{p} \right):{f_2}\left( {{p_1}^{\left( j \right)},{p_2}^{\left( j \right)}} \right) + \frac{{\partial {f_2}}}{{\partial {p_1}}}\left( {{p_1}^{\left( j \right)},{p_2}^{\left( j \right)}} \right){\Delta p_1} + \frac{{\partial {f_2}}}{{\partial {p_2}}}\left( {{p_1}^{\left( j \right)},{p_2}^{\left( j \right)}} \right){\Delta p_2} = 0
\end{array} \right.
\end{aligned}
\label{equ:set2_HDUA}
\end{equation}
\end{figure*}

Using Algorithm \ref{alg:LNmethod_HDUA}, we finally find $p_1$ and $p_2$ to optimize the sum rate $r$. Taking account of the inequality constraints on the time and energy variables, we are able to find a local optimal solution of time and energy allocations as
\begin{equation}
\begin{aligned}
\left\{ \begin{aligned}
t_1 = \left[\frac{{p_2 - \frac{E}{K}}}{{p_2 - p_1}}\right]^+,\\
t_2 = \left[\frac{{\frac{E}{K} - p_1}}{{p_2 - p_1}}\right]^+,\\
\varepsilon _1 = \left[p_1\frac{{p_2 - \frac{E}{K}}}{{p_2 - p_1}}\right]^+,\\
\varepsilon _2 = \left[p_2\frac{{\frac{E}{K} - p_1}}{{p_2 - p_1}}\right]^+,
\end{aligned} \right.
\end{aligned}
\end{equation}
where $[x]^+=\max(x,0)$.

\subsection{ASR of HD Bidirectional Link with NUPA}
\label{subsection3.2}
\subsubsection{Problem Formulation}
In this subsection, we
optimize the ASR of an HD bidirectional link with NUPA. In this case, different powers are allocated on different sub-carriers. The achievable rates of the forward and backward links in the HD case are shown in \eqref{eq_HD-NUPA_r1} and \eqref{eq_HD-NUPA_r2}, respectively, where ${{\bar{\varepsilon }}_i}=[\varepsilon_i[1],\varepsilon_i[2],...,\varepsilon_i[K]]^{\rm{T}}$. Consequently, the sum rate of the whole system can be written as
\begin{equation}
r\left( {{{\bar{\varepsilon }}_1},{{\bar{\varepsilon }}_2},{t_1},{t_2}} \right) = {r_1}\left( {{{\bar{\varepsilon }}_1},{t_1}} \right)+{r_2}\left( {{{\bar{\varepsilon }}_2},{t_2}} \right).
\end{equation}

\begin{figure*}
\begin{equation}\label{eq_HD-NUPA_r1}
\begin{aligned}
{r_1}\left( {{{\bar{\varepsilon }}_1},{t_1}} \right) = \frac{{{t_1}}}{K}\sum\limits_{k = 1}^K {{{\log }_2}\left( {1 + \frac{{{\varepsilon _1}\left[ k \right]{\gamma _E}{{\left| {{h_{21}}\left[ k \right]} \right|}^2}/{t_1}}}{{{\varepsilon _1}\left[ k \right]{{\left| {{h_{21}}\left[ k \right]} \right|}^2}/{t_1} + \left( {{\gamma _E} + 1} \right){N_2}}}} \right)},
\end{aligned}
\end{equation}
\end{figure*}
\begin{figure*}
\begin{equation}\label{eq_HD-NUPA_r2}
\begin{aligned}
{r_2}\left( {{{\bar{\varepsilon }}_2},{t_2}} \right) = \frac{{{t_2}}}{K}\sum\limits_{k = 1}^K {{{\log }_2}\left( {1 + \frac{{{\varepsilon _2}\left[ k \right]{\gamma _E}{{\left| {{h_{12}}\left[ k \right]} \right|}^2}/{t_2}}}{{{\varepsilon _2}\left[ k \right]{{\left| {{h_{12}}\left[ k \right]} \right|}^2}/{t_2} + \left( {{\gamma _E} + 1} \right){N_1}}}} \right)},
\end{aligned}
\end{equation}
\end{figure*}

To achieve the maximum ASR, the following optimization problem
can be formulated:
\begin{equation} \label{eq_NUPA_HD_Opt1}
\begin{aligned}
\mathop {{\rm{minimize}}}_{{{\bar{\varepsilon }}_1},~{{\bar{\varepsilon }}_2},~{t_1},~{t_2}} {\rm{ }}\quad &-r\left( {{{\bar{\varepsilon }}_1},{{\bar{\varepsilon }}_2},{t_1},{t_2}} \right),\\
{\rm{subject\ to\quad }}&{t_1} + {t_2} = 1,\\
&\sum\limits_{k = 1}^K {{\varepsilon _1}\left[ k \right]}  + \sum_{k = 1}^K {{\varepsilon _2}\left[ k \right]}  = E,\\
&-\bar\varepsilon _1 \preceq {\bar{0}},~-\bar\varepsilon _2 \preceq {\bar{0}},\\
&-{t_1} \leq 0,~-{t_2} \leq 0.
\end{aligned}
\end{equation}

\subsubsection{Solution of the Problem}
It is clear that this problem is again non-convex. Similar to the UPA case in the previous subsection, we are interested in finding a suboptimal solution with three steps.

\emph{Step 1:} Let us first deal with a partial Lagrangian function, which is given by
\begin{equation}
\begin{aligned}
 &L\left( {{{\bar{\varepsilon }}_1},{{\bar{\varepsilon }}_2},{t_1},{t_2},\lambda ,v} \right)= -r\left( {{{\bar{\varepsilon }}_1},{{\bar{\varepsilon }}_2},{t_1},{t_2}} \right)  \\
  &+\lambda \left( {\sum\limits_{k = 1}^K {\left( {{\varepsilon _1}\left[ k \right] + {\varepsilon _2}\left[ k \right]} \right)}  - E} \right)   + \nu \left( {{t_1} + {t_2} - 1} \right), \\
\end{aligned}
\end{equation}
where $v$ is the Lagrange multiplier of time constraint $t_1 + t_2 =1$ and $\lambda $ is the Lagrange multiplier of energy constraint $\sum_{k = 1}^K {\left( {{\varepsilon _1}\left[ k \right] + {\varepsilon _2}\left[ k \right]} \right)}  - E = 0$.

\emph{Step 2:} We formulate a new problem, which is equivalent to the original problem in \eqref{eq_NUPA_HD_Opt1}, to minimize the partial Lagrangian function as follows.
\begin{equation} \label{eq_NUPA_HD_Opt1_partial}
\begin{aligned}
\mathop {{\rm{minimize}}}_{{{\bar{\varepsilon }}_1},~{{\bar{\varepsilon }}_2},~{t_1},~{t_2},\lambda,v} {\rm{ }}\quad &L\left( {{{\bar{\varepsilon }}_1},{{\bar{\varepsilon }}_2},{t_1},{t_2},\lambda ,v} \right),\\
{\rm{subject\ to\quad }}&{t_1} + {t_2} = 1,\\
&\sum\limits_{k = 1}^K {{\varepsilon _1}\left[ k \right]}  + \sum_{k = 1}^K {{\varepsilon _2}\left[ k \right]}  = E,\\
&-\bar\varepsilon _1 \preceq {\bar{0}},~-\bar\varepsilon _2 \preceq {\bar{0}},\\
&-{t_1} \leq 0,~-{t_2} \leq 0.
\end{aligned}
\end{equation}

\emph{Step 3:} We solve \eqref{eq_NUPA_HD_Opt1_partial} by formulating an equation set and proposing a numerical method to solve the equation set.

At a local optima, we have
\begin{equation} \label{eq_NUPA_HD_Opt2}
\begin{aligned}
\frac{{\partial L}}{{\partial {t _1}}} =\frac{{\partial L}}{{\partial {t _2}}} = \frac{{\partial L}}{{\partial {\varepsilon _1}\left[ k \right]}} =\frac{{\partial L}}{{\partial {\varepsilon _2}\left[ k \right]}} = 0,~k=1,2,...,K.\\
\end{aligned}
\end{equation}

Consequently, we obtain equations
\begin{equation} \label{eq_constraint_T}
\frac{{\partial {r_1}}}{{\partial {t_1}}} = \frac{{\partial {r_2}}}{{\partial {t_2}}} = v,
\end{equation}
and
\begin{equation}
\frac{{\partial r}}{{\partial {\varepsilon _1}\left[ k \right]}} = \frac{{\partial r}}{{\partial {\varepsilon _2}\left[ k \right]}} = \lambda,~k=1,2,...,K,
\end{equation}
where $\frac{{\partial r}}{{\partial {\varepsilon _1}\left[ k \right]}}$ and $\frac{{\partial r}}{{\partial {\varepsilon _2}\left[ k \right]}}$ are computed as \eqref{equ:deriveofr_HDAA1} and \eqref{equ:deriveofr_HDAA2}, respectively, while $\frac{{\partial {r_1}}}{{\partial {t_1}}}$ and $\frac{{\partial {r_2}}}{{\partial {t_2}}}$ are computed as \eqref{equ:deriveofr_HDAA3} and \eqref{equ:deriveofr_HDAA4}, respectively.

The remaining task is to solve these equations. Note that these equations cannot be divided into small equation arrays, because all the energy variables are coupled in $\frac{{\partial {r_1}}}{{\partial {t_1}}}$ and $\frac{{\partial {r_1}}}{{\partial {t_2}}}$. Although it may be feasible to use the Newton's method to solve these equations, a linear equation array analogous to \eqref{equ:set2_HDUA} needs to be solved in each iteration. When $K$ is big, the approach will be impractical. Fortunately, as we can see from \eqref{equ:deriveofr_HDAA1} to \eqref{equ:deriveofr_HDAA4}, given $t_1$, the other variables can be found without solving the large-scale equation array. Hence, we propose a method of grid search on $t_1$ within $[0, 1]$. For each value of $t_1$ within $[0, 1]$, we have $t_2=1-t_1$, and we can further obtain ${\varepsilon _1}\left[ k \right]$ and ${\varepsilon _2}\left[ k \right]$ by the following process.

\begin{figure*}[!t]
\begin{equation}
\frac{{\partial r}}{{\partial {\varepsilon _1}\left[ k \right]}} = \frac{{{t_1}}}{K}\frac{{{\gamma _E}{{\left| {{h_{21}}\left[ k \right]} \right|}^2}{N_2}}}{{\left[ {{\varepsilon _1}\left[ k \right]{{\left| {{h_{21}}\left[ k \right]} \right|}^2} + \left( {{\gamma _E} + 1} \right){N_2}{t_1}} \right]\left[ {{\varepsilon _1}\left[ k \right]{{\left| {{h_{21}}\left[ k \right]} \right|}^2} + {N_2}{t_1}} \right]}} = \lambda
\label{equ:deriveofr_HDAA1}
\end{equation}
\end{figure*}
\begin{figure*}[!t]
\begin{equation}
\frac{{\partial r}}{{\partial {\varepsilon _2}\left[ k \right]}} = \frac{{1 - {t_1}}}{K}\frac{{{\gamma _E}{{\left| {{h_{12}}\left[ k \right]} \right|}^2}{N_1}}}{{\left[ {{\varepsilon _2}\left[ k \right]{{\left| {{h_{12}}\left[ k \right]} \right|}^2} + \left( {{\gamma _E} + 1} \right){N_1}\left( {1 - {t_1}} \right)} \right]\left[ {{\varepsilon _2}\left[ k \right]{{\left| {{h_{12}}\left[ k \right]} \right|}^2} + {N_1}\left( {1 - {t_1}} \right)} \right]}} = \lambda
\label{equ:deriveofr_HDAA2}
\end{equation}
\end{figure*}
\begin{figure*}[!t]
\begin{equation}
\frac{{\partial {r_1}}}{{\partial {t_1}}} = \frac{1}{K}{{\sum_{k = 1}^K {{{\log }_2}\frac{{\left( {{\varepsilon _1}\left[ k \right]{{\left| {{h_{21}}\left[ k \right]} \right|}^2} + {N_2}{t_1}} \right)\left( {{\gamma _E} + 1} \right)}}{{{\varepsilon _1}\left[ k \right]{{\left| {{h_{21}}\left[ k \right]} \right|}^2} + \left( {{\gamma _E} + 1} \right){N_2}{t_1}}}} }} - \frac{{t_1}}{{K\ln 2}}{{\sum_{k = 1}^K {\frac{{{\varepsilon _1}\left[ k \right]{{\left| {{h_{21}}\left[ k \right]} \right|}^2}{\gamma _E}{N_2}}}{{\left( {{\varepsilon _1}\left[ k \right]{{\left| {{h_{21}}\left[ k \right]} \right|}^2} + {N_2}{t_1}} \right)\left( {{\varepsilon _1}\left[ k \right]{{\left| {{h_{21}}\left[ k \right]} \right|}^2} + \left( {{\gamma _E} + 1} \right){N_2}{t_1}} \right)}}} }}
\label{equ:deriveofr_HDAA3}
\end{equation}
\end{figure*}

\begin{figure*}[!t]
\begin{equation}
\frac{{\partial {r_2}}}{{\partial {t_2}}} = \frac{1}{K}{{\sum_{k = 1}^K {{{\log }_2}\frac{{\left( {{\varepsilon _2}\left[ k \right]{{\left| {{h_{12}}\left[ k \right]} \right|}^2} + {N_1}{t_2}} \right)\left( {{\gamma _E} + 1} \right)}}{{{\varepsilon _2}\left[ k \right]{{\left| {{h_{12}}\left[ k \right]} \right|}^2} + \left( {{\gamma _E} + 1} \right){N_1}{t_2}}}} }} - \frac{t_2}{{K\ln 2}}{{\sum_{k = 1}^K {\frac{{{\varepsilon _2}\left[ k \right]{{\left| {{h_{12}}\left[ k \right]} \right|}^2}{\gamma _E}{N_1}}}{{\left( {{\varepsilon _2}\left[ k \right]{{\left| {{h_{12}}\left[ k \right]} \right|}^2} + {N_1}{t_2}} \right)\left( {{\varepsilon _2}\left[ k \right]{{\left| {{h_{12}}\left[ k \right]} \right|}^2} + \left( {{\gamma _E} + 1} \right){N_1}{t_2}} \right)}}} }}
\label{equ:deriveofr_HDAA4}
\end{equation}
\end{figure*}

The derivatives in \eqref{equ:deriveofr_HDAA1} and \eqref{equ:deriveofr_HDAA2} can be rewritten into quadratic equations, and thus we can solve ${\varepsilon _1}\left[ k \right]$ and ${\varepsilon _2}\left[ k \right]$ provided that $t_1$ and $t_2$ are given. For instance, regarding to
${\varepsilon _1}\left[ k \right]$, we have
\begin{equation}
a{\varepsilon _1}{\left[ k \right]^2} + b{\varepsilon _1}\left[ k \right] + c = 0,
\label{equ:abc_equation}
\end{equation}
where
\begin{equation}
\begin{aligned}
a = & {\left| {{h_{21}}\left[ k \right]} \right|^4},\\
b = & {\left| {{h_{21}}\left[ k \right]} \right|^2}{N_2}{t_1}\left( {{\gamma _E} + 2} \right),\\
c = & \left( {{\gamma _E} + 1} \right)N_2^2t_1^2 - \frac{{{\gamma _E}{{\left| {{h_{21}}\left[ k \right]} \right|}^2}{N_2}{t_1}}}{{K\lambda }}.
\end{aligned}
\end{equation}
It is easy to verify that this quadratic equation has surely two solutions. When $c<0$, there are a positive and negative solution, respectively, while when $c\geq 0$, there are two non-positive solutions. Considering that ${\varepsilon _1}\left[ k \right]\geq 0$, the bigger one, or zero if the bigger one is negative, is chosen as the solution for ${\varepsilon _1}\left[ k \right]$, i.e.,
\begin{equation} \label{eq_solution_epsi}
{\varepsilon _1}{\left[ k \right]} = \left[{ {\frac{{ - b + \sqrt {{b^2} - 4ac} }}{{2a}}} }\right ]^+.
\end{equation}
${\varepsilon _2}\left[ k \right]$ can also be found by
a similar approach. It can be observed that when $t_1$ is given, ${\varepsilon _1}\left[ k \right]$ and ${\varepsilon _2}\left[ k \right]$ are functions of $\lambda$, and they monotonically increase as $\lambda$ decreases according to \eqref{equ:deriveofr_HDAA1} and \eqref{equ:deriveofr_HDAA2}. Thus, $\lambda $ can be found by the bisection method to meet the energy constraint, as shown in Algorithm \ref{alg:bisection1}.

\begin{algorithm} \caption{Computation of the Lagrange Multiplier $\lambda$ via the Bisection Method.}
\label{alg:bisection1}
\begin{algorithmic}
\STATE \textbf{1) Initialization:}
\begin{quote}

Find an arbitrary positive multiplier ${\lambda _{\rm{left}}}$ that meets $\sum_{k = 1}^K {\left( {{\varepsilon _1}\left[ k \right] + {\varepsilon _2}\left[ k \right]} \right)}  - E > 0$.

Find an arbitrary positive multiplier ${\lambda _{\rm{right}}}$ that meets $\sum_{k = 1}^K {\left( {{\varepsilon _1}\left[ k \right] + {\varepsilon _2}\left[ k \right]} \right)}  - E < 0$.

Set a positive small variable $\delta$ as the energy allocation error, i.e., $\delta \ll 1$.

\end{quote}

\STATE \textbf{2) Iteration:}
\begin{quote}
\textbf{For} $j=1,2,...,$
\begin{adjustwidth}{0.5 cm}{0 cm}

Set ${\lambda ^{\left( j \right)}} = ({{{\lambda _{\rm{left}}} + {\lambda _{\rm{right}}}}})/{2}$, and compute ${\varepsilon _i}{\left[ k \right]}$ as \eqref{eq_solution_epsi}, $i=1,2;~k=1,2,...,K$.

\textbf{If} $|\sum_{k = 1}^K {\left( {{\varepsilon _1}\left[ k \right] + {\varepsilon _2}\left[ k \right]} \right)} - E|<\delta$
\begin{adjustwidth}{0.5 cm}{0 cm}
$\lambda  = {\lambda ^{\left( j \right)}}$;

Break;
\end{adjustwidth}
\textbf{ELSE}
\begin{adjustwidth}{0.5 cm}{0 cm}
\textbf{If} $\sum_{k = 1}^K {\left( {{\varepsilon _1}\left[ k \right] + {\varepsilon _2}\left[ k \right]} \right)}  - E > 0$
\begin{adjustwidth}{0.5 cm}{0 cm}
Update as ${\lambda _{\rm{left}}} = {\lambda ^{\left( j \right)}}$.
\end{adjustwidth}
\textbf{ELSE}
\begin{adjustwidth}{0.5 cm}{0 cm}
Update as ${\lambda _{\rm{right}}} = {\lambda ^{\left( j \right)}}$.
\end{adjustwidth}
\end{adjustwidth}
\end{adjustwidth}
\end{quote}

\STATE \textbf{3) Result:}
\begin{quote}
Normalize ${\varepsilon _1}[k]$ and ${\varepsilon _2}[k]$ such that $\sum_{k = 1}^K {\left( {{\varepsilon _1}\left[ k \right] + {\varepsilon _2}\left[ k \right]} \right)}  - E = 0$.

Return the Lagrange multiplier $\lambda $.
\end{quote}

\end{algorithmic}
\end{algorithm}

Now we have obtained $t_2$, ${\varepsilon _1}\left[ k \right]$ and ${\varepsilon _2}\left[ k \right]$ by assuming that $t_1$ is given. The problem
that remains unsolved is to find the value of $t_1$ within $[0,1]$ based on the condition in \eqref{eq_constraint_T}, which can be obtained by
a grid search method within $[0,1]$, as illustrated
in Algorithm \ref{alg:HD-AA_time}.

\begin{algorithm} \caption{Grid Search Method for $t_1$.}
\label{alg:HD-AA_time}
\begin{algorithmic}
\STATE \textbf{1) Initialization:}
\begin{quote}
Set a constraint error $\delta  \ll 1$.

Set the step length $\xi  \ll 1$.
\end{quote}

\STATE \textbf{2) Iteration:}
\begin{quote}
\textbf{For} ${t_1} = 0,\xi ,2\xi ,...,1$
\begin{adjustwidth}{0.5 cm}{0 cm}

Compute ${\varepsilon}_1[k]$,  ${\varepsilon}_2[k]$ and $\lambda$ utilizing Algorithm \ref{alg:bisection1};


\textbf{If} $|\frac{{\partial {r_1}}}{{\partial {t_1}}} - \frac{{\partial {r_2}}}{{\partial {t_2}}}|<\delta$

\begin{adjustwidth}{0.5 cm}{0 cm}
Break;
\end{adjustwidth}
\end{adjustwidth}
\end{quote}

\STATE \textbf{3) Result:}
\begin{quote}
Return the time allocation ${t_1}$, ${t_2} = 1 - {t_1}$, as well as the power allocations $\varepsilon_1[k]|_{t_1}$ and $\varepsilon_2[k]|_{t_1}$.
\end{quote}

\end{algorithmic}
\end{algorithm}

\subsection{ASR of FD Bidirectional Link with UPA}
\subsubsection{Problem Formulation}
When utilizing
the UPA strategy over sub-carriers on FD bidirectional link, the consumed energy on the $k$-th sub-carrier becomes
${\varepsilon _1}\left[ k \right] = {\varepsilon _1}$ for the forward link and ${\varepsilon _2}\left[ k \right] = {\varepsilon _2}$ for backward link. It is noted that there is no time allocation for the FD transmission. Therefore, the forward and backward rates can be
derived as \eqref{equ:r1_FDUA} and \eqref{equ:r2_FDUA}, respectively. Accordingly, the ASR of the whole transmission system becomes
\begin{equation}
r\left( {{\varepsilon _1},{\varepsilon _2}} \right) = {r_1}\left( {{\varepsilon _1},{\varepsilon _2}} \right) + {r_2}\left( {{\varepsilon _1},{\varepsilon _2}} \right).
\end{equation}
To achieve the maximum ASR, we must solve
the optimization problem as follows:
\begin{equation}
\begin{aligned} \label{eq_optproblem3}
\mathop {{\rm{maximize}}}\limits_{{\varepsilon _1},{\varepsilon _2}} {\rm{ }}\quad &r\left( {{\varepsilon _1},{\varepsilon _2}} \right),\\
{\rm{subject\ to\quad }}&K\left( {{\varepsilon _1} + {\varepsilon _2}} \right) = E,\\
&{\varepsilon _1} \ge 0,{\varepsilon _2} \ge 0.
\end{aligned}
\end{equation}

\begin{figure*}[!t]
\begin{equation}
{r_1}\left( {{\varepsilon _1},{\varepsilon _2}} \right) = \frac{1}{K}\sum\limits_{k = 1}^K {{{\log }_2}\left( {1 + \frac{{{\varepsilon _1}{\gamma _E}{{\left| {{h_{21}}\left[ k \right]} \right|}^2}}}{{{\varepsilon _1}{{\left| {{h_{21}}\left[ k \right]} \right|}^2} + {\varepsilon _2}{\gamma _E}{{\left| {{h_{22}}\left[ k \right]{\beta _2}\left[ k \right]} \right|}^2} + {\varepsilon _2}{{\left| {{h_{22}}\left[ k \right]} \right|}^2} + \left( {{\gamma _E} + 1} \right){N_2}}}} \right)}
\label{equ:r1_FDUA}
\end{equation}
\begin{equation}
{r_2}\left( {{\varepsilon _1},{\varepsilon _2}} \right) = \frac{1}{K}\sum\limits_{k = 1}^K {{{\log }_2}\left( {1 + \frac{{{\varepsilon _2}{\gamma _E}{{\left| {{h_{12}}\left[ k \right]} \right|}^2}}}{{{\varepsilon _2}{{\left| {{h_{12}}\left[ k \right]} \right|}^2} + {\varepsilon _1}{\gamma _E}{{\left| {{h_{11}}\left[ k \right]{\beta _1}\left[ k \right]} \right|}^2} + {\varepsilon _1}{{\left| {{h_{11}}\left[ k \right]} \right|}^2} + \left( {{\gamma _E} + 1} \right){N_1}}}} \right)}
\label{equ:r2_FDUA}
\end{equation}
\end{figure*}

\subsubsection{Solution of the Problem}
Apparently, \eqref{eq_optproblem3} is also non-concave when observing from the expressions. However, under typical conditions, i.e., the thermal noise power is much lower than the EVM noise power, the SINR is much higher than 0 dB, and the channel is symmetric, an optimal solution of \eqref{eq_optproblem3} can be (approximately) obtained (see Appendix A), which is $\varepsilon_1=\varepsilon_2=E/2$. Under other conditions, however, an optimal solution of \eqref{eq_optproblem3} is difficult to obtain; thus, numerical methods can be considered instead to find a suboptimal solution. Since there are only two variables for this problem, the Newton's method with backtracking line search, which is referred to \cite[Chapter 9.5]{boyd2004convex}, can be adopted. Details are not presented for conciseness. It is noted that if the typical conditions are satisfied, the solution found by the Newton's method is optimal; otherwise it may be not. This is verified in Figs. \ref{fig:flatchange}, \ref{fig:selectchange} and \ref{fig:select2change}, where we can see that the achievable rates of the two nodes are the same at the searched points in Figs. \ref{fig:flatchange} and \ref{fig:selectchange} under a symmetric channel, while different in Fig. \ref{fig:select2change} under an asymmetric channel.


\subsection{ASR of FD Bidirectional Link with NUPA}
\subsubsection{Problem Formulation}
When utilizing adaptive power allocation strategy over sub-carriers, the forward and backward rates of the FD bidirectional link are
given as in \eqref{equ:r1_FDAA} and \eqref{equ:r2_FDAA}. Accordingly, the ASR of the whole transmission system becomes
\begin{equation}
r\left( {{{{\bf{\bar \varepsilon }}}_1},{{{\bf{\bar \varepsilon }}}_2}} \right) = {r_1}\left( {{{{\bf{\bar \varepsilon }}}_1},{{{\bf{\bar \varepsilon }}}_2}} \right) + {r_2}\left( {{{{\bf{\bar \varepsilon }}}_1},{{{\bf{\bar \varepsilon }}}_2}} \right).
\end{equation}

\begin{figure*}[!t]
\begin{equation}
{r_1}\left( {{{\bar{\varepsilon }}_1},{{\bar{\varepsilon }}_2}} \right) = \frac{1}{K}\sum\limits_{k = 1}^K {{{\log }_2}\left( {1 + \frac{{{\varepsilon _1}\left[ {\rm{k}} \right]{\gamma _E}{{\left| {{h_{21}}\left[ k \right]} \right|}^2}}}{{{\varepsilon _1}\left[ {\rm{k}} \right]{{\left| {{h_{21}}\left[ k \right]} \right|}^2} + {\varepsilon _2}\left[ {\rm{k}} \right]{\gamma _E}{{\left| {{h_{22}}\left[ k \right]{\beta _2}\left[ k \right]} \right|}^2} + {\varepsilon _2}\left[ {\rm{k}} \right]{{\left| {{h_{22}}\left[ k \right]} \right|}^2} + \left( {{\gamma _E} + 1} \right){N_2}}}} \right)}
\label{equ:r1_FDAA}
\end{equation}
\begin{equation}
{r_2}\left( {{{\bar{\varepsilon }}_1},{{\bar{\varepsilon }}_2}} \right) = \frac{1}{K}\sum\limits_{k = 1}^K {{{\log }_2}\left( {1 + \frac{{{\varepsilon _2}\left[ {\rm{k}} \right]{\gamma _E}{{\left| {{h_{12}}\left[ k \right]} \right|}^2}}}{{{\varepsilon _2}\left[ {\rm{k}} \right]{{\left| {{h_{12}}\left[ k \right]} \right|}^2} + {\varepsilon _1}\left[ {\rm{k}} \right]{\gamma _E}{{\left| {{h_{11}}\left[ k \right]{\beta _1}\left[ k \right]} \right|}^2} + {\varepsilon _1}\left[ {\rm{k}} \right]{{\left| {{h_{11}}\left[ k \right]} \right|}^2} + \left( {{\gamma _E} + 1} \right){N_1}}}} \right)}
\label{equ:r2_FDAA}
\end{equation}
\end{figure*}

To achieve the maximum ASR, we have
the following optimization problem:
\begin{equation} \label{eq_FD-NUPA_Problem}
\begin{aligned}
\mathop {{\rm{maximize}}}\limits_{{{\bar{\varepsilon }}_1},{{\bar{\varepsilon }}_2}} {\rm{ }}\quad &r\left( {{{\bar{\varepsilon }}_1},{{\bar{\varepsilon }}_2}} \right), \\
{\rm{subject\ to\quad }}&\sum\limits_{k = 1}^K {\left( {{\varepsilon _1}\left[ k \right] + {\varepsilon _2}\left[ k \right]} \right)}  = E,\\
&\bar\varepsilon _1 \succeq {\bar{0}},~\bar\varepsilon _2 \succeq {\bar{0}}.
\end{aligned}
\end{equation}

\subsubsection{Solution of the Problem}
Again, \eqref{eq_FD-NUPA_Problem} is apparently non-concave when observing from the expressions. However, with similar proof in Appendix A, we can prove that under typical conditions that the thermal noise power is much lower than the EVM noise power, the SINR is much higher than 0 dB, and the channel is symmetric, an optimal solution of \eqref{eq_FD-NUPA_Problem} can be (approximately) found. In particular, under the typical conditions $\varepsilon_1[k]=\varepsilon_2[k]$ can be derived first at the optima, and then they can be determined by using the water-filling approach. Under other conditions, however, an optimal solution is difficult to find. Thus, we prefer numerical approaches instead to find a suboptimal solution. Different from the ASR optimization of FD with UPA in \eqref{eq_optproblem3}, where the Newton's method with backtracking line search is feasible because there are only two variables, the problem in \eqref{eq_FD-NUPA_Problem} has $2K$ variables. Hence, when $K$ is big, the Newton's method with backtracking line search is impractical, because a large-scale equation array needs to be solved in each iteration. In fact, this is just the case encountered in problem \eqref{eq_NUPA_HD_Opt1}. However, there is no time allocation in the problem \eqref{eq_FD-NUPA_Problem}.

Similar to the UPA case, we are interested in finding a suboptimal solution with three steps.

\emph{Step 1:} Let us first deal with a partial Lagrangian function, which is given by
\begin{equation}
L\left( {{{\bar{\varepsilon }}_1},{{\bar{\varepsilon }}_2},\lambda} \right)=r\left( {{{\bar{\varepsilon }}_1},{{\bar{\varepsilon }}_2}} \right) - \lambda \left( {\sum\limits_{k = 1}^K {\left( {{\varepsilon _1}\left[ k \right]{\rm{ + }}{\varepsilon _2}\left[ k \right]} \right)}  - E} \right).
\end{equation}

\emph{Step 2:} We formulate a new problem, which is equivalent to the original problem in \eqref{eq_FD-NUPA_Problem}, to maximize the partial Lagrangian function as follows.
\begin{equation} \label{eq_FD-NUPA1_Problem}
\begin{aligned}
\mathop {{\rm{maximize}}}_{{{\bar{\varepsilon }}_1},~{{\bar{\varepsilon }}_2},\lambda} {\rm{ }}\quad &L\left( {{{\bar{\varepsilon }}_1},{{\bar{\varepsilon }}_2},\lambda} \right),\\
{\rm{subject\ to\quad }}&\sum\limits_{k = 1}^K {\left( {{\varepsilon _1}\left[ k \right] + {\varepsilon _2}\left[ k \right]} \right)}  = E,\\
&\bar\varepsilon _1 \succeq {\bar{0}},~\bar\varepsilon _2 \succeq {\bar{0}}.
\end{aligned}
\end{equation}

\emph{Step 3:} We solve \eqref{eq_FD-NUPA1_Problem} by formulating an equation set and proposing a numerical method to solve the equation set.

At a local optima, we have
\begin{equation} \label{eq_FD_NUPA_KKT}
\begin{aligned}
&\frac{{\partial r}}{{\partial {\varepsilon _1}\left[ k \right]}}= \frac{{\partial r}}{{\partial {\varepsilon _2}\left[ k \right]}} = \lambda,~k=1,2,...,K,\\
&\sum\limits_{k = 1}^K {\left( {{\varepsilon _1}\left[ k \right]{\rm{ + }}{\varepsilon _2}\left[ k \right]} \right)}  - E = 0.\\
\end{aligned}
\end{equation}

The remaining task is to solve these equations. Recall that in the solution of problem \eqref{eq_NUPA_HD_Opt1}, where there are also plenty of variables, we proposed the approach of grid search on $t_1$, because all the other variables can be conveniently found when $t_1$ is given. We can adopt a similar approach here, i.e., we propose the grid search on $\lambda$ from 0 with a step length $\xi$ until the energy constraint is satisfied, which is similar to Algorithm \ref{alg:HD-AA_time}, because when $\lambda$ is given, ${\varepsilon _1}[k]$ and ${\varepsilon _2}[k]$ can be solved according to \eqref{eq_FD_NUPA_KKT}. It is emphasized that although there are $2K$ variables to be solved when $\lambda$ is given, these variables are paired in terms of carrier index $k$, i.e., ${\varepsilon _1}[k]$ and ${\varepsilon _2}[k]$ are paired together, and they are independent to ${\varepsilon _1}[j]$ and ${\varepsilon _2}[j]$ when $j\neq k$. Hence, when $\lambda$ is given, ${\varepsilon _1}[k]$ and ${\varepsilon _2}[k]$ can be obtained by exploiting the Newton's method with backtracking line search, as shown in Algorithm \ref{alg:LNmethod_HDUA}, to solve the two-variable equation array $\frac{{\partial r}}{{\partial {\varepsilon _1}\left[ k \right]}}= \frac{{\partial r}}{{\partial {\varepsilon _2}\left[ k \right]}} = \lambda$. It is noted that if the typical conditions are satisfied, the solution found by the proposed method is optimal; otherwise it may be not. This is verified in Figs. \ref{fig:flatchange}, \ref{fig:selectchange} and \ref{fig:select2change}, where we can see that the achievable rates of the two nodes are the same at the searched points in Figs. \ref{fig:flatchange} and \ref{fig:selectchange} under a symmetric channel, while different in Fig. \ref{fig:select2change} under an asymmetric channel.


\subsection{Discussions on the Optimality and Complexity}

We have formulated the problems of ASR maximization and derived numerical approaches to them for the four transmission strategies in this section, namely HD-UPA, HD-NUPA, FD-UPA, and FD-NUPA. For HD-UPA, HD-NUPA and FD-NUPA,
by dealing with a partial Lagrangian function, we first formulate an equivalent new problem. Then we establish an equation array with time and energy variables and adopt appropriate numerical methods to solve the equation array. For FD-UPA, we directly adopt a numerical method to optimize the ASR, because there are only two variables. The involved numerical search algorithms are Newton's method with backtracking line search, bisection search and grid search. Among them, the bisection search and grid search are naturally convergent, while the convergence of the Newton's method with backtracking line search has been proven \cite[Chapters 9.2 and 9.5]{boyd2004convex}.

Since the proposed solutions are numerical, the achieved performance depends on the step length of the search algorithms. It is clear that the performance is better when the step length is smaller, whereas the number of iterations will be larger. Given sufficiently small step lengths, as the ASR optimization problems for HD-UPA and HD-NUPA are non-convex, the proposed solutions to them are suboptimal. For FD-UPA and FD-NUPA, the solutions can be seen optimal under the typical conditions, while suboptimal under other conditions.

Next, we evaluate the complexities of the proposed approaches, where complexity refers to the number of iterations or searches here. The exhaustive grid search is considered as a comparative scheme. It is emphasized that although the search complexity of the interior-point method is also low, as the Newton's method is also adopted in the search process of the interior-point method \cite[Chapters 10 and 11]{boyd2004convex}, a large-scale linear equation array is needed to be solved in each iteration in the interior-point method, while not needed in the proposed method.

\begin{itemize}
  \item Regarding the problem in \eqref{eq_HD-UPA_Problem} for HD-UPA, there are four variables with two equality constraints. Thus, the number of independent variables is in fact 2. The exhaustive grid search on the 2 independent variables has a complexity of $\mathcal{O}(N^2)$, where $1/N$ can be seen as the step length of the grid search. We propose the Lagrange duality approach and exploit the Newton's method with backtracking line search to obtain a suboptimal solution. According to \cite[Chapter 9.5]{boyd2004convex}, the required number of iterations is almost constant, i.e., $\mathcal{O}(1)$.
  \item Regarding the problem in \eqref{eq_NUPA_HD_Opt1} for HD-NUPA, the number of independent variables is $2K$. Thus, the complexity of an exhaustive grid search scheme is $\mathcal{O}(N^{2K})$. We propose to perform grid search on $t_1$ as shown in Algorithm \ref{alg:HD-AA_time}, and in each iteration, a bisection search is embedded. Hence, the complexity can be expressed as $\mathcal{O}(N)\mathcal{O}(\log_2 N)$.
  \item Regarding the problem in \eqref{eq_optproblem3} for FD-UPA, the number of independent variables is $1$. Thus, the complexity of an exhaustive grid search scheme is $\mathcal{O}(N)$. In contrast, the proposed Newton's method with backtracking line search has a complexity of $\mathcal{O}(1)$ \cite[Chapter 9.5]{boyd2004convex}.
  \item Regarding the problem in \eqref{eq_FD-NUPA_Problem} for FD-UPA, the number of independent variables is $2K-1$. Thus, the complexity of an exhaustive grid search scheme is $\mathcal{O}(N^{2K-1})$. We propose to perform grid search on $\lambda$, and in each iteration, the Newton's method with backtracking line search is launched $K$ times. Hence, the complexity is $K\mathcal{O}(1)\mathcal{O}(N)$.
\end{itemize}

The results in this subsection are summarized in Table \ref{tab:complexity_comp}.

\begin{table}[tbp]
\centering  
\caption{Optimality and Complexity of the Proposed Approaches, where $1/N$ can be seen as the step length, $K$ is the number of sub-carriers, EGS represents Exhaustive Grid Search.}
\label{tab:complexity_comp}
\begin{tabular}{cccc}  
\hline
Strategies & Optimality & Complexity &\begin{tabular}{c} Complexity \\of EGS \end{tabular}\\ \hline  
HD-UPA & Suboptimal & $\mathcal{O}(1)$ & $\mathcal{O}(N^2)$ \\         
HD-NUPA & Suboptimal & $\mathcal{O}(N)\mathcal{O}(\log_2 N)$ & $\mathcal{O}(N^{2K})$ \\        
FD-UPA & Optimal/Suboptimal & $\mathcal{O}(1)$ & $\mathcal{O}(N)$ \\
FD-NUPA & Optimal/Suboptimal & $K\mathcal{O}(1)\mathcal{O}(N)$ & $\mathcal{O}(N^{2K-1})$\\
\hline
\end{tabular}
\end{table}



\section{Numerical Results}
\label{section4}
 In this section, we present the numerical results on the ASR performance of the four transmission strategies, i.e., HD-UPA, HD-NUPA, FD-UPA, and FD-NUPA, taking account of three different channels, namely symmetric frequency-flat channel, symmetric frequency-selective channel, and asymmetric frequency-selective channel. Moreover, under the asymmetric frequency-selective channel, which is the most general one in the three channel models, we compared the ASR performance achieved by the proposed schemes with that by Li's method in \cite{Wei2014} and the exhaustive search, respectively. Some of the system parameters are listed in Table \ref{tab:basic}.

\begin{table}[tbp]
\centering  
\caption{Part of Involved System Parameters.}
\label{tab:basic}
\begin{tabular}{lccc}  
\hline
Parameters & Values & Units \\ \hline  
Signal bandwidth & 10 & MHz \\         
Carrier frequency & 2 & GHz \\        
Antenna gain & 0 & dB \\
Thermal noise power density at receiver & -174 & dBm/Hz \\
Receiver noise figure & 10 & dB \\
EVM level & -30 & dBc \\
Number of sub-carriers & 64 &  \\
Baseband interference cancellation & -40 & dB \\
Distance of two transceivers & 30 & m \\
\hline
\end{tabular}
\end{table}

\subsection{Symmetric Frequency-Flat Channel}
In the
frequency-flat scenario, it is assumed that all channels have the same gain over all sub-carriers. The channel gains between the two transceivers in the communication link are found by utilizing the outdoor line-of-sight (LOS) Path-Loss (PL) model \cite{Wei2014}:
\begin{equation}
PL = 103.8 + 20.9{\log _{10}}\frac{d}{{1000}}(dB),
\end{equation}
where $PL$ (dB) is the path loss, $d$ (m) is the distance between the two transceivers. Besides, in the performance evaluations,
the unit total time is assumed (thus,
the maximum transmission power constraint
becomes the energy constraint).
All the ASRs are achieved by using the optimization approaches proposed in Section III.

\begin{figure}[t]
\begin{center}
  \includegraphics[width=\figwidth cm]{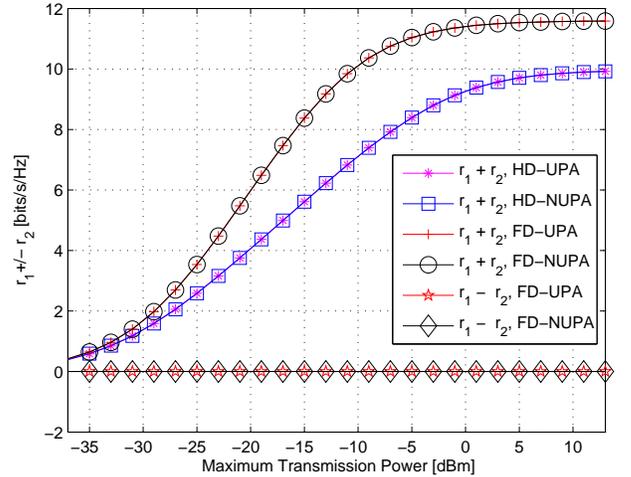}
  \caption{ASRs w.r.t. maximum transmission power in symmetric frequency-flat channel. $h_{11}=h_{22}=-60$ dB.}
  \label{fig:flatchange}
\end{center}
\end{figure}

\begin{figure}[t]
\begin{center}
  \includegraphics[width=\figwidth cm]{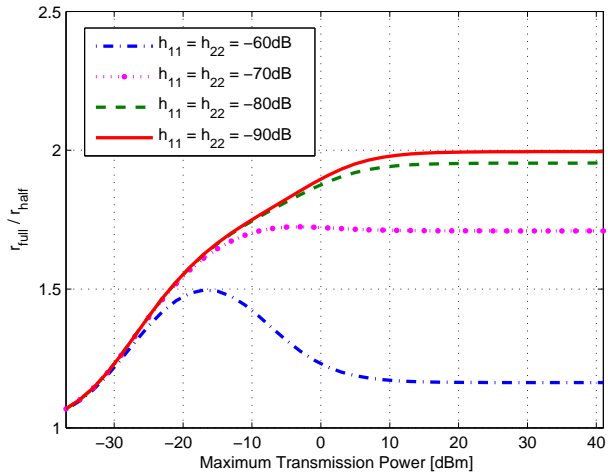}
  \caption{Ratio of transmission rates of FD link and that of HD link in symmetric frequency-flat channel.}
  \label{fig:flatratio}
\end{center}
\end{figure}

Fig. \ref{fig:flatchange} shows the ASRs of the four transmission strategies
w.r.t. to the
maximum transmission power in the case of $h_{11}=h_{22}=-60$ dB under symmetric frequency-flat channel, where $r_1+r_2$ represents ASR,
while $r_1-r_2$ is the difference of achievable rate between the two nodes,
and $h_{11}$ and $h_{22}$ are the power attenuations of the SI channels in the two nodes. It is found that the ASRs of the four strategies increase with the maximum transmission power. However, the increasing
rate becomes slower as the maximum transmission power becomes higher, and eventually the ASRs achieve a maximal value and do not further increase.
This phenomenon is due to the SI and EVM noise whose powers are proportional to the transmission power, and thus even with high transmission power, the effective SINR, as well as the ASR, will not be further improved. Meanwhile, we find that in symmetric frequency-flat channel, UPA achieves the same performance as NUPA for both FD and HD transmissions, which are in accordance with our expectation. Besides, for FD transmission the backward achievable rate is almost the same as the forward achievable rate to reach the maximal ASR.

From Fig. \ref{fig:flatchange}, it can also be observed that FD transmission outperforms HD transmission. Basically, we can expect that FD transmission achieves double ASR over HD transmission, provided that SI can be completely cancelled. However, in practice where residual SI exists, the superiority of FD transmission may be affected. Fig. \ref{fig:flatratio} shows the ratio of ASRs of FD transmission to that of HD transmission under the same system conditions in symmetric frequency-flat channel. It is found that when the residual SI is low, e.g., $h_{11}=h_{22}=-90$ or $-80$ dB in Fig. \ref{fig:flatratio}, the ratio basically
increases with the maximum transmission power, and finally reaches a maximal value close to 2. However, when the residual SI is high, e.g., $h_{11}=h_{22}=-60$ or $-70$ dB in Fig. \ref{fig:flatratio}, the ratio has
a peak as the maximum transmission power increases.
Since the residual SI becomes significant when the transmission power is high enough, the EVM noise and SI will become dominant over thermal noise,
which can offset the increase of the rate of the FD transmission, and thus makes the ratio decreases after the peak.

\subsection{Symmetric frequency-selective channel}
\label{subsection4.2}
In this subsection we consider a symmetric frequency-selective channel. To establish the channel models, firstly we choose two time-domain multipath models for the signal transmission channel and the SI channel, respectively. The same as that in \cite{Wei2014}, the transmission channel is modeled based on ITU outdoor channel model A with multipath components (MPC) as shown in Table \ref{tab:select}.
While the SI channel is generated from a 4-tap time-domain multipath model as
\begin{equation}
\begin{aligned}
h_{\rm{SI}}\left[ t \right] = {A}\frac{{\exp \left( { - t} \right)}}{{\sum\limits_{t = 0}^3 {\exp \left( { - t} \right)} }},~t = 0,1,2,3,
\end{aligned}
\end{equation}
where $A$ is the power attenuation of the SI channel. Since the channel is symmetric, it is assumed that ${h_{12}}\left[ k \right] = {h_{21}}\left[ k \right]$ and ${h_{11}}\left[ k \right] = {h_{22}}\left[ k \right]$.

\begin{table}[tbp]
\centering  
\caption{Multipath Components for Symmetric Frequency-Selective Transmission Channel.}
\label{tab:select}
\begin{tabular}{lccc}  
\hline
Tap & Relative delay (ns) & Average power (dB) \\ \hline  
1 & 0 & 0 \\
2 & 300 & -1 \\
3 & 700 & -9 \\
4 & 1100 & -10 \\
5 & 1700 & -15 \\
6 & 2500 & -20 \\
\hline
\end{tabular}
\end{table}

\begin{figure}[t]
\begin{center}
  \includegraphics[width=\figwidth cm]{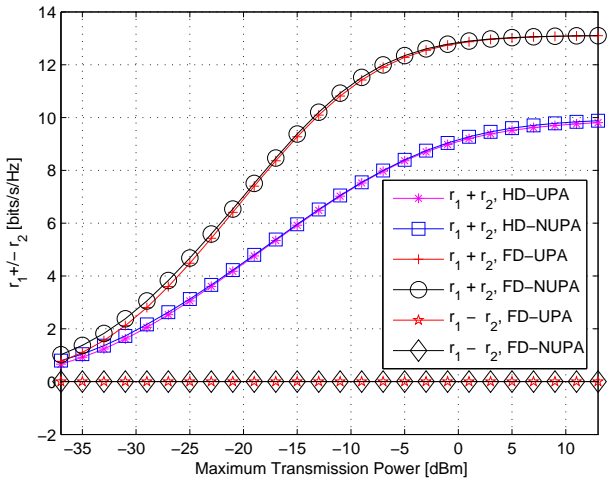}
  \caption{ASRs w.r.t. maximum transmission power in symmetric frequency-selective channel. $h_{11}=h_{22}=-60$ dB.}
  \label{fig:selectchange}
\end{center}
\end{figure}

Fig. \ref{fig:selectchange} shows the ASRs of the four transmission strategies w.r.t. to
the maximum transmission power in the case of $h_{11}=h_{22}=-60$ dB under symmetric frequency-selective channels, where $h_{11}$ and $h_{22}$ are the power attenuations of the SI channels in the two nodes. Similar to that in the frequency-flat channel, the ASRs of the four strategies increase
with the maximum transmission power, but the increasing
rate becomes slower and eventually the ASRs achieve a maximal value and do not further increase. Besides, for FD transmission the backward achievable rate is the same as the forward achievable rate to reach the maximal ASR, which are in accordance with our expectation, because the frequency-selective channel is also symmetric here. In addition, we can observe that NUPA does provide a better ASR performance than UPA for both FD and HD transmissions, but the superiority is not significant, especially in high-transmission-power regime. The superiority
results from the adaptive power allocation
(water-filling analogous operations in the optimization process). However, due to the presence of EVM noise, which reduces the SINR difference on each sub-carrier, the adaptive power allocation cannot achieve a significant superiority.

\begin{figure}[t]
\begin{center}
  \includegraphics[width=\figwidth cm]{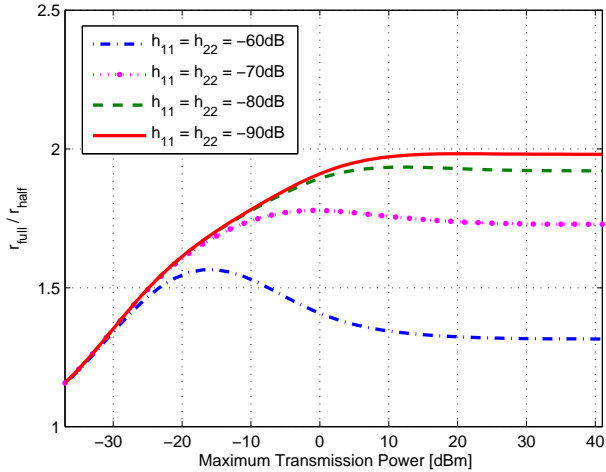}
  \caption{Ratio of transmission rates of FD link and that of HD link with NUPA in symmetric frequency-select channel.}
  \label{fig:selectratio}
\end{center}
\end{figure}

 From Fig. \ref{fig:selectchange}, it can also be observed that FD transmission outperforms HD transmission. Fig. \ref{fig:selectratio} shows the ratio of ASRs of FD transmission to that of HD transmission with NUPA under the same system conditions in symmetric frequency-selective channel. Similarly, it is found that when the residual SI is low, the ratio basically increases
with the maximum transmission power, and finally reach a maximal value close to 2. However, when the residual SI is high, the ratio
has a $\cap$-shape with a peak.

\subsection{Asymmetric frequency-selective channel}
In this subsection, we consider an asymmetric frequency-selective channel without channel reciprocity. Two asymmetric multipath models are adopted with their tap coefficients shown in Table \ref{tab:select2_1} for the transmission channel and Table \ref{tab:select2_2} for the SI channel, respectively.

\begin{table}[tbp]
\centering  
\caption{Multipath Components for asymmetric Frequency-Selective Transmission Channel.}
\label{tab:select2_1}
\begin{tabular}{lccc}  
\hline
Tap & Link 1 to 2 & Link 2 to 1 \\ \hline  
1 & 9.9863e2 & j1.4921e3 \\
2 & 2.6934e2 & 1.1503e3 \\
3 & j3.3458e2 & 0.8853e3 \\
4 & 3.1862e2 & 1.1284e3 \\
5 & j2.1856e2 & 0.1637e3 \\
6 & 0.9111e2 & j0.4007e3 \\
\hline
\end{tabular}
\end{table}

\begin{table}[tbp]
\centering  
\caption{Multipath Components for asymmetric Frequency-Selective SI Channel.}
\label{tab:select2_2}
\begin{tabular}{lccc}  
\hline
Tap & Link 1 to 1 & Link 2 to 2 \\ \hline  
1 & 1.3103e2 & 1.3712e2 \\
2 & j1.6827e2 & 0.3585e2 \\
3 & 1.3241e2 & j0.4396e2 \\
4 & 1.0621e2 & j0.2212e2 \\
\hline
\end{tabular}
\end{table}

\begin{figure}[t]
\begin{center}
  \includegraphics[width=\figwidth cm]{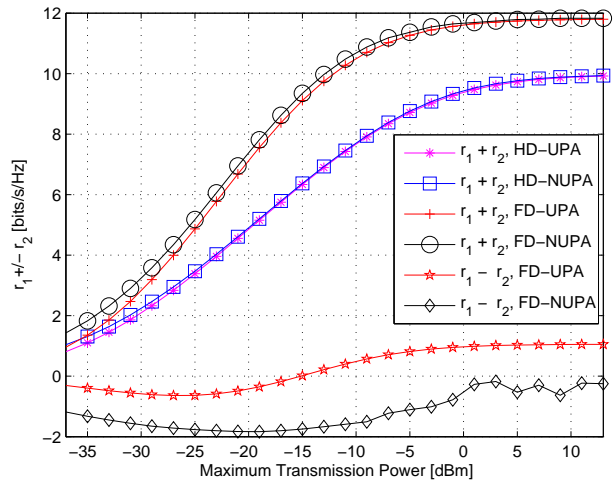}
  \caption{ASRs w.r.t. maximum transmission power in asymmetric frequency-selective channel. $h_{11}=h_{22}=-60$ dB.}
  \label{fig:select2change}
\end{center}\vspace{-0.1 in}
\end{figure}

 Fig. \ref{fig:select2change} shows the ASRs of the four transmission strategies w.r.t. the maximum transmission power in the case of $h_{11}=h_{22}=-60$ dB under asymmetric frequency-selective channels. Comparing Fig. \ref{fig:select2change} with Fig. \ref{fig:selectchange}, we can find that the asymmetric channel setting does not affect much on
the ASR performance. In particular, the same results can be observed from Fig. \ref{fig:select2change} as those from Fig. \ref{fig:selectchange}, except that for FD transmission the backward achievable rate is no longer equal to the forward achievable rate to reach the optimal ASR due to the asymmetry of the channels.

\begin{figure}[t]
\begin{center}
  \includegraphics[width=\figwidth cm]{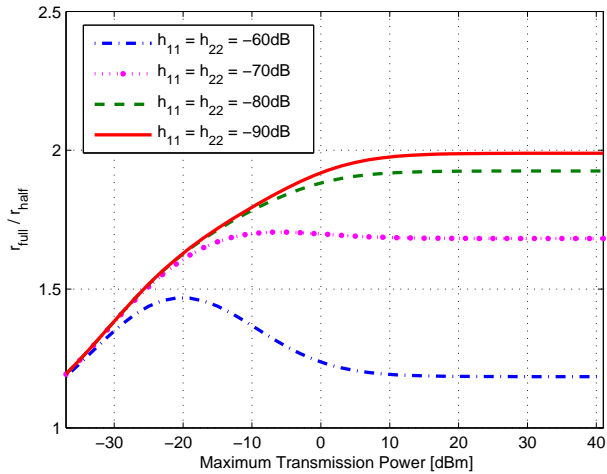}
  \caption{Ratio of transmission rates of FD link and that of HD link with NUPA in asymmetric frequency-select channel. $h_{11}=h_{22}=-60$ dB.}
  \label{fig:select2ratio}
\end{center}
\end{figure}

 Fig. \ref{fig:select2ratio} shows the ratio of ASR of FD transmission
to that of HD transmission with NUPA under the same system conditions in asymmetric frequency-selective channel. The same results are observed as those from Fig. \ref{fig:selectratio}.

\begin{figure}[t]
\begin{center}
  \includegraphics[width=\figwidth cm]{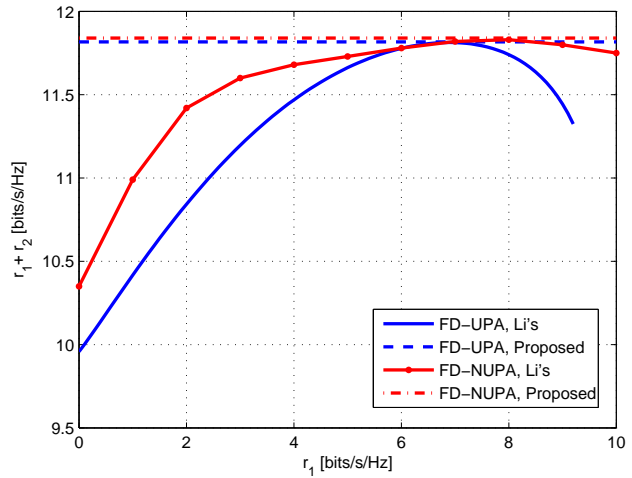}
  \caption{ASR comparison between the proposed schemes and Li's method under asymmetric frequency-selective channel. $h_{11}=h_{22}=-60$ dB, and the maximum transmission power is $5$ dBm. The $r_1$-axis is for Li's method, not for the proposed methods, because the proposed methods do no have the constraint of $r_1$.}
  \label{fig_oricompare}
\end{center}
\end{figure}

Now we have evaluated the ASR performance achieved by the proposed methods for the four transmission strategies under the three different channel models. Next, we want to compare the ASR performance of the proposed methods with that of Li's method proposed in \cite{Wei2014}, where the backward rate ($r_2$) is optimized with a given forward rate $r_1$ and individual energy constraints. Since only one-end rate instead of ASR was optimized and individual energy constraints instead of total energy constraint was adopted in \cite{Wei2014}, it can be expected that the proposed methods basically achieve better ASR performance. Fig. \ref{fig_oricompare} shows the ASR comparison results for the FD transmission under the asymmetric frequency-selective channel, where the maximum transmission power is $5$ dBm. It can be found that for both FD-UPA and FD-NUPA with Li's methods, there are peaks of ASR for $r_1\in[0,~10]$ bits/s/Hz, which is in accordance with the results in \cite{Wei2014} (Fig. 7 therein). Besides, the ASR with the proposed methods, which does not depend on $r_1$, is basically better than that with Li's method, and the superiority depends on $r_1$. The peaks of the ASR curves with Li's method are close to the ASR achieved by the proposed methods, which means that the individual energy constraints in the simulation are good to achieve high ASR. In other words, the individual energy constraints happen to be close to the optimal energy allocations of Node 1 and Node 2 with the proposed methods. However, due to the individual energy constraints, the peaks are basically no higher than the ASR lines with the proposed methods, where total energy constraint is adopted. With other individual energy constraints, the peaks may be lower than the ASR lines with the proposed methods. For instance, if the individual energy constraints are $E$ for Node 1 and $0$ for Node 2 in Li's method, the achieved ASR will be equal to $r_1$, which is far lower than the ASRs with the proposed methods. Similar results can be also observed for HD transmission.

\begin{figure}[t]
\begin{center}
  \includegraphics[width=\figwidth cm]{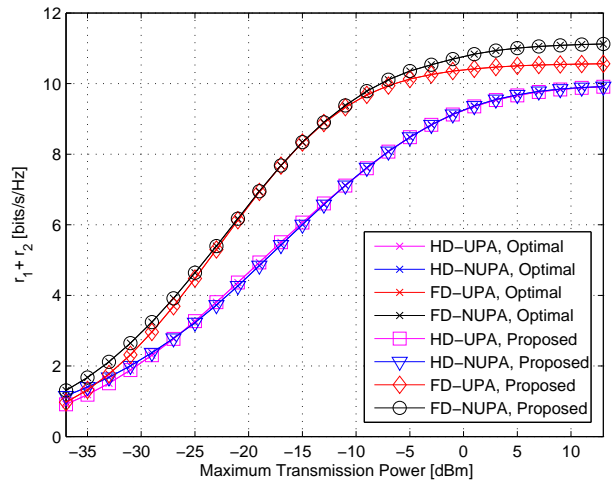}
  \caption{ASR comparison between the proposed schemes and the exhaustive grid search scheme under asymmetric frequency-selective channel. $K=4$.}
  \label{fig_comparison}
\end{center}
\end{figure}

Moreover, in order to evaluate the optimality of the proposed methods, we compared the ASR performance of the proposed methods with that of the exhaustive grid search scheme under the asymmetric frequency-selective channel in Fig. \ref{fig_comparison}, where $K$ is set to 4 for computation feasibility. The exhaustive grid search method can provide the optimal results provided that the step length for each variable is sufficiently small. From this figure it can be observed that the achieved ASRs of these four solutions are almost the same as those obtained by the optimal ones. These results demonstrate the effectiveness of the proposed methods.

\section{Conclusions}
\label{section5}
In this paper we have studied the ASR performances
of HD and FD transmissions with UPA and NUPA strategies, where the non-ideality of practical transceivers, modeled as the EVM noise, is taken into account and the typical three-stage SI cancellation process is adopted.
Four optimization problems have been formulated to maximize the ASRs
of the four transmission strategies.
A low-complexity approach, which first deals with a partial Lagrangian function without taking into account the inequality constraints and formulate an equivalent problem of the original problem, and then solves the new problem by exploiting appropriate numerical methods,
has been proposed to find solutions to the fourth formulated problems, which
has been shown to achieve near-optimal performances. For FD transmissions optimal solutions can be obtained under the typical conditions, i.e., the thermal noise power is far less than the EVM noise power, SINR is much higher than 0 dB, and the channel is symmetric.

From simulation results, we can have the following observations:
i) Basically, FD transmission outperforms HD transmission in
terms of
ASR performance for
symmetric/asymmetric frequency-flat/selective channels, and the superiority depends on the residual SI. The ratio of ASR of FD to
that of HD can reach to the limit, 2,
given that
the residual SI is low enough. (ii) Both the ASRs of FD and HD transmissions become improved as the maximum transmission power increases, but due to the existence of EVM noise and SI, the increasing rate
gradually slows down, and the ASRs eventually become
saturated. (iii) Owing to the existence of EVM and SI, the adaptive power allocation, i.e., NUPA, does not
perform significantly better than UPA. (iv) In the case that the ASR is maximized, the achievable rates of the forward and backward links for FD transmission are almost the same in symmetric channels, while evidently different in asymmetric channels.

\appendices

\section{An Optimal Solution of Problem \eqref{eq_optproblem3}}
In this section we prove that an optimal solution of \eqref{eq_optproblem3} can be (approximately) obtained with typical conditions, i.e., the thermal noise power is much lower than the EVM noise power, the SINR is much higher than 0 dB, and the channel is symmetric. Let $A_k=|h_{21}[k]|^2=|h_{12}[k]|^2$, $B_k=\gamma_{E} |h_{22}[k]\beta_2[k]|^2+|h_{22}[k]|^2=\gamma_{E} |h_{11}[k]\beta_1[k]|^2+|h_{11}[k]|^2$. Under the typical conditions, the thermal noise can be neglected and we have
\begin{equation}
\begin{aligned}
r&\approx\frac{1}{K}\sum_{k=1}^{K} \log_2\left(\frac{\varepsilon_1\gamma_E A_k}{\varepsilon_1A_k+\varepsilon_2 B_k}\frac{\varepsilon_2\gamma_E A_k}{\varepsilon_2A_k+\varepsilon_1 B_k} \right)\\
&=\frac{1}{K}\sum_{k=1}^{K} \log_2\left(\frac{\gamma_E A_k}{A_k+x^{-1} B_k}\frac{\gamma_E A_k}{A_k+x B_k} \right),
\end{aligned}
\end{equation}
where $x=\varepsilon_1/\varepsilon_2$. We further obtain
\begin{equation}
\frac{\partial r}{\partial x}=-\frac{1}{K\ln 2}\sum_{k=1}^{K}\frac{A_kB_k(1-x^{-2})}{A_k^2+A_kB_k(x+x^{-1})+B_k^2}.
\end{equation}

At the optimal point we have $\frac{\partial r}{\partial x}=0\Rightarrow x=1$. Thus, the optimal solution is $\varepsilon_1=\varepsilon_2=E/2$ under the typical conditions. It should be noteworthy that this result is obtained by neglecting the noise power and the constant 1 within the $\log$ function of $r$ in \eqref{eq_optproblem3} under the typical conditions. Under other conditions the conclusion may not hold.


\end{document}